\documentclass[preprints,article,accept,moreauthors,pdftex]{mdpi} 

\firstpage{1} 
\pubvolume{1}
\issuenum{1}
\articlenumber{0}
\pubyear{2021}
\copyrightyear{2020}
\datereceived{} 
\dateaccepted{} 
\datepublished{} 
\hreflink{https://doi.org/} % If needed use \linebreak
\usepackage{graphicx}
\usepackage{epsfig}
\usepackage{rotating}
\usepackage{amssymb}
\usepackage{subfigure}
\usepackage{dsfont}
\usepackage{psfrag}
\usepackage{mathtools}
\usepackage{amsmath}
\usepackage{euscript}
\usepackage{array}
\usepackage{mathrsfs}
\usepackage{bbold}
\usepackage{bm}
\usepackage{bbm}
\usepackage{multirow}
\usepackage{dcolumn}
\usepackage{xcolor}

\newcommand{\beq}{\begin{equation}}
\newcommand{\eeq}{\end{equation}}
\newcommand{\beqs}{\begin{eqnarray}}
\newcommand{\eeqs}{\end{eqnarray}}

\newcommand{\Tr}{{\rm Tr}}

\def\cl{{\cal L}}
\def\cm{{\cal M}}

\def\cs{{\cal S}}

\Title{Dilaton Effective Field Theory }

\TitleCitation{Dilaton Effective Field Theory }

\Author{Thomas Appelquist, $^{1}$\orcidA{}, 
James Ingoldby $^{2}$\orcidB{}, 
and Maurizio Piai $^{3}$\orcidC{}}

\AuthorNames{Thomas Appelquist, James Ingoldby, and Maurizio Piai}

\AuthorCitation{
Appelquist, T.; Ingoldby, J.;  Piai, M.}
\address[5]{%
$^1$ \quad  Department of Physics, Sloane Laboratory, Yale University, New Haven, Connecticut 06520, USA\\
$^2$ \quad Abdus Salam International Centre for Theoretical Physics, Strada Costiera 11, 34151, Trieste, Italy\\
$^3$ \quad Department of Physics, Faculty of Science and Engineering,
Swansea University (Singleton Park Campus), Singleton Park, SA2 8PP Swansea, Wales, United Kingdom}
\abstract{ 
	We review and extend recent studies of dilaton effective field theory (dEFT) which provide a framework for the description of the Higgs boson as a composite structure. We first describe the dEFT as applied to lattice data for a class of gauge theories with near-conformal infrared behavior. It includes the dilaton associated with the spontaneous breaking of (approximate) scale invariance, and a set of  pseudo-Nambu-Goldstone bosons (pNGBs) associated with the spontaneous breaking of an (approximate) internal global symmetry. The theory contains two small symmetry-breaking parameters. We display the leading-order (LO) Lagrangian, and review its fit to lattice data for the $SU(3)$ gauge theory with $N_f = 8$ Dirac fermions in the fundamental representation. We then develop power-counting rules to identify the corrections emerging at next-to-leading order (NLO) in the dEFT action. We list the NLO operators that appear and provide estimates for the coefficients. We comment on implications for composite-Higgs model building. %We list the NLO operators that appear, provide estimates for the corresponding coefficients, and suggest some future avenues of research.
}

\keyword{lattice gauge theory, physics beyond the standard model}

\begin{document}

%%%%%%%%%%%%%%%%%%%%%%%%%%%%%%%%%%%%%%%%%%
%\setcounter{section}{-1} %% No need to uncomment. 
%%%
%%%UNCOMMENT THESE FOUR LINES IF YOU WANT A TOC
%%%\end{paracol}
%%%\newpage
%%%\tableofcontents
%%%\newpage
%%%

%\newpage
%%%%%%%%%%%%%%%%%%%%%%%%sec1%%%%%%%%%%%%%%%%%%%%%%%%%%%%%%%%%%%%%%%%%%%%%%%%%%%%%%
\section{Introduction}
\label{sec:introduction}
%%%%%%%%%%%%%%%%%%%%%%%%sec1%%%%%%%%%%%%%%%%%%%%%%%%%%%%%%%%%%%%%%%%%%%%%%%%%%%%%%

It has long been thought that the dilaton, the neutral Nambu-Goldstone boson (NGB) arising from the 
spontaneous breaking of scale invariance, might play a role in fundamental 
physics---see, e.g., Refs~\cite{Migdal:1982jp,Coleman:1985rnk}. 
The idea is intriguing yet elusive. If an approximate symmetry under scale transformations sets in over some energy range, 
and if the forces are such that this symmetry is not respected by the ground state (vacuum) of the system, 
then the appearance of an approximate, light dilaton would seem natural.

It is easy to realize this possibility at the classical level, there being no better example than the Higgs potential of the standard model (SM) with its minimum at a vacuum expectation value (VEV) $v_W > 0$ of the Higgs field. The Higgs particle becomes lighter as the self--coupling strength is reduced with fixed $v_W$. In this limit, the explicit breaking of scale invariance by the Higgs mass becomes smaller, and the breaking dominantly spontaneous, due to the VEV $v_W$. The Higgs particle can then be viewed as an approximate dilaton at the classical level. 

The dilaton idea becomes more subtle at the quantum level. At either level, it makes sense only if the explicit 
breaking is relatively small, so that there is an approximate scale invariance (dilatation symmetry) to 
be broken spontaneously. In a quantum field theory, 
explicit breaking arises not only from the fixed dimensionful parameters in the Lagrangian, but also through the renormalization process. For example, the quantum corrections to the Higgs potential can lead to large contributions to the Higgs mass, requiring fine tuning to maintain its lightness. Yet, its interpretation as a dilaton has striking implications for the standard model as well as its extensions~\cite{Goldberger:2008zz}.

In a gauge theory like quantum chromodynamics (QCD), renormalization leads to a confinement scale $\Lambda$, explicitly breaking dilatation symmetry. Approximate scale invariance sets in only at higher energies, while the vacuum structure and composite-particle spectrum are determined at scales of order $\Lambda$ itself. There is no reason to expect the appearance of an approximate dilaton in the composite spectrum. On the other hand, as the number of light fermions in a gauge theory is increased, the running of the gauge coupling slows, and it has been speculated that an approximate dilatation symmetry can develop, 
to be broken spontaneously at scales relevant to bound-state 
formation~\cite{Leung:1985sn,Bardeen:1985sm,Yamawaki:1985zg}. 
This idea, suggesting the presence of an approximate dilaton, has been supported by recent lattice studies. Gauge theories in this class both confine and have near--conformal behavior. 

Lattice studies of $SU(3)$ gauge theories with $N_f=8$  flavors of fundamental (Dirac) fermions~\cite{Aoki:2014oha,Appelquist:2016viq,Aoki:2016wnc,Gasbarro:2017fmi,Appelquist:2018yqe,LatticeStrongDynamicsLSD:2021gmp,Hasenfratz:2022qan}, as well as $N_f = 2$ flavors of symmetric 2-index (Dirac) fermions (sextets)~\cite{Fodor:2012ty,
	Fodor:2015vwa,
	Fodor:2016pls,Fodor:2017nlp,Fodor:2019vmw,Fodor:2020niv},
have reported evidence for the presence of a surprisingly light flavor--singlet scalar particle in the accessible range of fermion masses.  
Motivated by the possibility that such a particle might be an approximate dilaton, we analyzed the lattice data in terms of an effective field theory (EFT) framework that extends the field content of a conventional chiral Lagrangian~\cite{Appelquist:2017wcg,Appelquist:2017vyy,Appelquist:2019lgk,Appelquist:2020bqj,Appelquist:2022qgl}. It includes a dilaton field $\chi$, together with the pseudo-Nambu-Goldstone-boson (pNGB) fields $\pi$ describing the other light composite particles revealed by the
lattice studies. 

Gauge theories that are near conformal are particularly interesting because dEFT can provide a low energy description of a light composite Higgs boson as an approximate dilaton~\cite{Goldberger:2008zz,Hong:2004td,Dietrich:2005jn,Hashimoto:2010nw,Appelquist:2010gy,Vecchi:2010gj,Chacko:2012sy,Bellazzini:2012vz,Bellazzini:2013fga,Abe:2012eu,Eichten:2012qb,Hernandez-Leon:2017kea}. They could also form the basis for a realistic composite Higgs model in which the Higgs boson is  an admixture of the dilaton state and one of the pNGBs~\cite{Appelquist:2020bqj,Appelquist:2022qgl}. In this context, electroweak quantum numbers must be assigned to the fermions of the gauge theory, and a coupling to the top quark must be included. Having an EFT description of the lightest degrees of freedom is then a valuable model building tool. In both cases, precision Higgs physics could reveal the composite nature of the Higgs boson, with dEFT providing a framework for this endeavor.

Here we revisit the dilaton-effective-field-theory (dEFT) description of the light particle spectrum of these gauge theories \cite{Appelquist:2017wcg,Appelquist:2017vyy,Appelquist:2019lgk,Appelquist:2020bqj,Appelquist:2022qgl}, which has also been examined in Refs.~\cite{Matsuzaki:2013eva,Golterman:2016lsd,Kasai:2016ifi,Hansen:2016fri,Golterman:2016cdd,Golterman:2018mfm,Cata:2019edh,Cata:2018wzl,Golterman:2020tdq,Golterman:2020utm}. In Section~\ref{sec:LO}, we summarize the underlying principles of the dEFT, describe the leading-order (LO) effective Lagrangian, and briefly recall the tree-level fit to lattice data carried out in Ref.~\cite{Appelquist:2019lgk}. In Section~\ref{sec:BLO}, we describe the dEFT more generally as a low-energy expansion, taking into account the effect of quantum loop corrections. Most importantly, we discuss the power-counting rules that are applied to improve the precision of the dEFT description, providing explicitly the form of the NLO Lagrangian. This extends work presented in Ref.~\cite{Cata:2019edh}. In Section~\ref{sec:discussion}, we summarize and comment on possible future applications.   

%%%%%%%%%%%%%%%%%%%%%%%%ack%%%%%%%%%%%%%%%%%%%%%%%%%%%%%%%%%%%%%%%%%%%%%%%%%%%%%%

%%%%%%%%%%%%%%%%%%%%%%%%sec1%%%%%%%%%%%%%%%%%%%%%%%%%%%%%%%%%%%%%%%%%%%%%%%%%%%%%%
\section{Leading Order (LO)}
\label{sec:LO}
%%%%%%%%%%%%%%%%%%%%%%%%sec1%%%%%%%%%%%%%%%%%%%%%%%%%%%%%%%%%%%%%%%%%%%%%%%%%%%%%%

To provide a low-energy description of explicit and spontaneous breaking of dilatation symmetry, we introduce a
scalar field $\chi$. It parametrizes approximately degenerate, but inequivalent, vacua, 
with dilatation symmetry spontaneously broken via a finite VEV  $\langle \chi \rangle = f_d$. 
The explicit breaking of dilatation symmetry yields a (small) mass $m_d$ for the dilaton,
the scalar particle associated with $\chi$.   

The dEFT also captures the spontaneous breaking of an approximate internal global symmetry group $\mathcal G$ 
to a subgroup $\mathcal H$. The  pNGBs are described  by the corresponding fields $\pi$. 
Their couplings are set by the decay constant $f_{\pi}$. 
A small  mass $m_{\pi}^2$ for the pNGBs  is present, as the global symmetry must be broken 
on the lattice. This explicit breaking also contributes to the full potential of the dilaton, as we shall see.

With ${\mathcal G} = SU(N_f)_L\times SU(N_f)_R$ and ${\mathcal H}=SU(N_f)_V$, the  Lagrangian density is  
\beqs 
{\cal L}_\text{LO}&=&\frac{1}{2}\partial_{\mu} \chi \partial^{\mu} \chi \,+\,{\cal L}_{K} \, +\,{\cal L}_M  \,-\,V_\Delta(\chi) \,.
\label{Eq:L} 
\eeqs 
The  kinetic term for the dilaton  takes canonical form. The pion kinetic terms 
\beqs 
{\cal L}_{K}&=&\frac{f_{\pi}^2}{4}\left(\frac{\chi}{f_{d}}\right)^2 \,\Tr\left[\partial_\mu \Sigma (\partial^{\mu} \Sigma)^{\dagger}\right]\,,
\label{Eq:Lpi} 
\eeqs 
are written in terms of the matrix-valued field $\Sigma = \exp[2i \pi/f_{\pi}]$. It
transforms as $\Sigma \rightarrow U_{L}\Sigma U_{R}^{\dagger}$ under the unitary transformations 
$U_{L,R} \in SU(N_{f})_{L,R}$, and it satisfies the nonlinear constraint $\Sigma \Sigma^{\dagger} = \mathbf{1}_{N_f}$. 

The dilaton potential $V_{\Delta}(\chi)$ takes the simple form 
\beqs
V_{\Delta}(\chi)&\equiv&\frac{m_d^2\chi^4}{4(4-\Delta)f_d^2}\left[1-\frac{4}{\Delta}\left(\frac{f_d}{\chi}\right)^{4-\Delta} \right]\,,
\label{Eq:Vdelta}
\eeqs
containing both a scale-invariant term ($\propto \chi^4$) and a scale-breaking term ($\propto \chi^{\Delta}$). 
The scaling parameter $\Delta$ is determined by a fit of the dEFT to lattice data. 
For any $\Delta$, the potential $V_{\Delta}(\chi)$ has a minimum at $\langle \chi \rangle = f_d$, with curvature $m_d^2$ 
at the minimum. In the limit $\Delta\rightarrow 4$ in which the deformation is near marginal, 
the potential smoothly approaches a functional form that includes a logarithm.

Explicit breaking of the global symmetry is described in the dEFT by 
\beqs
\mathcal{L}_M= \frac{m^2_\pi f^2_\pi}{4}\left(\frac{\chi}{f_d}\right)^y \Tr \left[\Sigma + \Sigma^\dagger\right]\,.
\label{Eq:LM}
\eeqs
The pion mass $m_{\pi}^2\equiv 2B_{\pi} m$ vanishes when the fermion mass in the underlying gauge theory, $m$, is set to zero. The quantity $B_\pi$ is a constant with dimensions of mass. The scaling dimension $y$ is determined by a fit of the dEFT to lattice data. It has been interpreted as the 
scaling dimension of the fermion bilinear condensate in the gauge theory in Ref.~\cite{Leung:1989hw}, and it has been suggested that near the 
edge of the conformal window $y$ approaches two~\cite{Cohen:1988sq,Ryttov:2017kmx}.
This interaction term also breaks scale invariance.

%%%%%%%%%%%%%%%%%%%%%%
\subsection{Mass Deformation and Scaling Properties}

In the presence of the mass deformation in Eq.~(\ref{Eq:LM}), 
and for $\langle\Sigma\rangle=\mathbf{1}_{N_f}$,
the complete dilaton potential  is given by
\beqs
W(\chi)= V_{\Delta}(\chi) \,-\,\frac{N_f m_{\pi}^2f_{\pi}^2}{2}\left(\frac{\chi}{f_d}\right)^y\,.
\label{Eq:Wpotential}
\eeqs
With $m_{\pi}^2 > 0$, $\chi$ develops a new VEV, $\langle \chi \rangle = F_d > f_d$, 
and a new curvature, (squared dilaton mass) $M_d^2$, 
near the minimum. The former is given by
\beqs
\left(\frac{F_d}{f_d}\right)^{4-y} \frac{1}{4-\Delta}\left[1 - \left(\frac{f_d}{F_d}\right)^{4-\Delta} \right]  = \frac{y N_{f} f_{\pi}^2 m_{\pi}^2}{2 f_{d}^2 m_{d}^2 } \,,
\label{Eq:Fdvar}
\eeqs
and the latter by
\beqs
\frac{M^2_d}{m^2_d}=\left(\frac{F_d}{f_d}\right)^2\frac{1}{4-\Delta}\left[4-y+(y-\Delta)\left(\frac{f_d}{F_d}\right)^{4-\Delta}\right]\,.
\label{Eq:Mdvar}
\eeqs

The dEFT leads to simple \emph{scaling relations} for pNGB decay constant and mass. They are
derived by normalizing the pNGB kinetic term in the vacuum, and read as follows:
\beqs
\frac{F_{\pi}^2}{f_{\pi}^2}= \left(\frac{F_{d}}{f_{d}}\right)^2\,,\qquad
\frac{M_{\pi}^2}{m_{\pi}^2}=\left( \frac{F_{d}}{f_{d}}\right)^{y-2}\,.\label{Eq:NGBscaling}
\eeqs
They are independent of the explicit form of the dilaton potential. 
They are directly useful in interpreting lattice data, 
leading for example to the relation $M_{\pi}^{2}F_{\pi}^{2-y}=Cm$, where 
$C=2B_{\pi}f_{\pi}^{2-y}$, which is used to measure $y$ from lattice data. When $\Delta$ is less than $4$ and $F_d \gg f_d$, Eq.~(\ref{Eq:Mdvar}) also simplifies into a simple scaling relation.

The dEFT Lagrangian density
can be recast in terms of the capital-letter quantities, 
a helpful rewrite for determination of the Feynman rules and interaction strengths. Taking $\chi = F_d + \bar{\chi}$, we have 
\beqs 
{\cal L}_K&=&\frac{F_{\pi}^2}{4}\left[ 1+ \frac{\bar{\chi}}{F_{d}}\right]^2 \,\Tr\left[\partial_\mu \Sigma (\partial^{\mu} \Sigma)^{\dagger}\right]\,,
\label{Eq:LpiR} 
\eeqs
and 
\beqs
{\cal L}_M&=&\frac{M_{\pi}^{2}F_{\pi}^2}{4}\left[1+ \frac{\bar{\chi}}{F_{d}}\right]^y\, \left[\Tr \left(\Sigma + \Sigma^{\dagger}\right) - 2N_f \right] \,,\label{Eq:LMR}
\eeqs
where $\Sigma = \exp\left[2i\Pi/F_\pi\right]$ and $\Pi$ are the canonically normalized pNGB fields. 
We have removed from Eq.~(\ref{Eq:LMR}) the piece that contributes to
the full dilaton potential $W(\chi)$, to avoid double counting. 
As an expansion in $\bar{\chi}/ F_d$, the potential takes the form
\beqs
W(\bar{\chi}) = \text{constant} + \frac{M_d^2}{2} \bar{\chi}^2 + \frac{\alpha}{3!} \frac{M_d^2}{F_d} \bar{\chi}^3 + \frac{\beta \, M_d^2}{4!\,F_d^2}\bar{\chi}^4\,+\,\cdots \,,
\label{Eq:Wexpansion}
\eeqs
where $\alpha$, $\beta$, $\dots$ are $O(1)$ dimensionless quantities depending on the dEFT parameters. When $F_d \gg f_d$,
for example, the parameter $\alpha$ varies between $3$ and $5$ as $\Delta$ varies from $2$ to the marginal-deformation case of $|4-\Delta|\ll1$ \cite{Goldberger:2008zz}.  

%%%%%%%%%%%%%%
\subsection{Fits to Lattice Data}
\label{subsec:Lat}

In Ref.~\cite{Appelquist:2019lgk}, we employed the LO expressions to perform a global, 
six-parameter fit to lattice data provided by the LSD collaboration for the $SU(3)$ gauge 
theory with $N_f = 8$ Dirac fermions in the fundamental representation. 
(Alternative analyses can be found, e.g., in Refs.~\cite{Fodor:2019vmw,Golterman:2020tdq,LatticeStrongDynamicsLSD:2021gmp}). As the scalar and pseudoscalar particles are much lighter than other composite states of the gauge theory for all available choices of the lattice parameters, it is sensible to describe them with our dEFT.
The data consists of values for $F_{\pi}^2$, $M_{\pi}^2$, and $M_d^2$ at five different 
values of the fermion mass $m$. We used the four dimensionless quantities $y$, 
$\Delta$, $f_\pi^2/f_d^2$ and $m_d^2/f_d^2$ as fit parameters, along 
with $f_{\pi}^2$ and $C$. All dimensionful quantities are expressed in units of the lattice spacing $a$. Central values and $1\sigma$ ranges for each of the six fit parameters can be found in Ref.~\cite{Appelquist:2019lgk}. The lattice data contains systematic uncertainties arising from finite volume and lattice discretization effects. It will be interesting to extend the dEFT to incorporate lattice artifacts systematically.

\begin{figure}[h]
	\begin{center}
		\includegraphics[height=4.8cm]{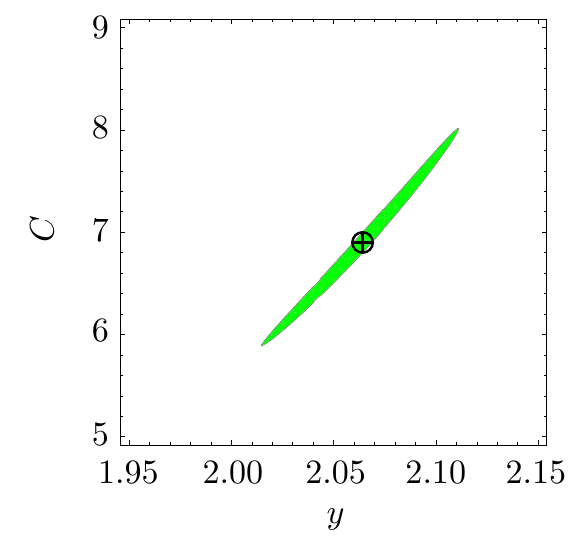}\qquad\qquad
		\includegraphics[height=4.8cm]{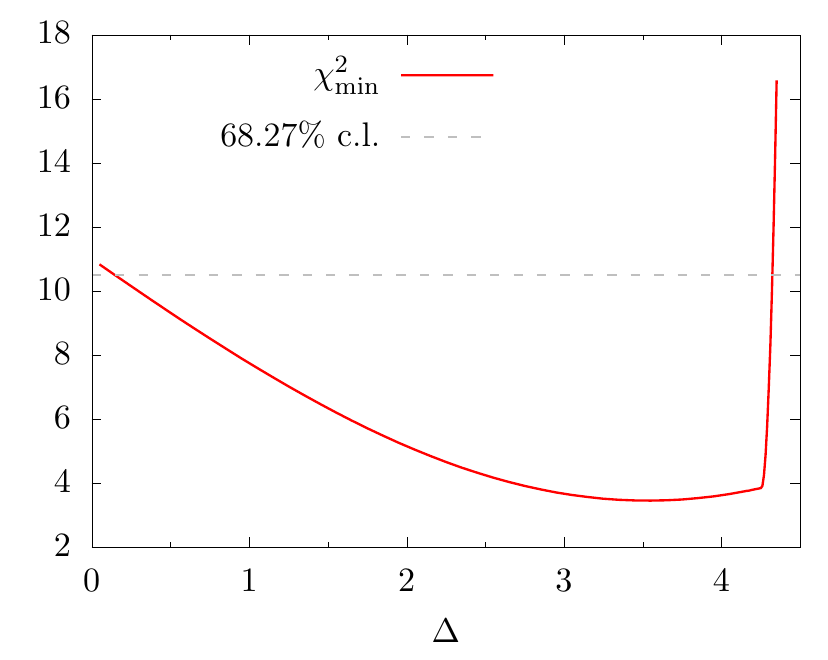}
		\caption{The 1$\sigma$ ranges for the scaling dimensions $y$ and $\Delta$, as defined in the main body of the
			paper, obtained in fits of LSD lattice measurements in the $SU(3)$ theory with $N_f=8$
			fundamental fermions~\cite{Appelquist:2019lgk}.}
		\label{Fig:yDeltafits}	
	\end{center}
\end{figure}

The relatively small uncertainties in the data for $F_{\pi}^2$ and $M_{\pi}^2$, 
along with the scaling relation $M_{\pi}^{2} F_{\pi}^{2-y} = C m$, 
allows for a relatively precise determination of $y$ and $C$. 
The correlated $1\sigma$ confidence ranges of these parameters is shown in 
the left panel of Fig.~\ref{Fig:yDeltafits},
in which the two plots are taken from Ref.~\cite{Appelquist:2019lgk}. 
At the $1\sigma$-equivalent confidence level, we found that
\beqs
y = 2.06 \pm 0.05 \, .
\eeqs
This range of values, if interpreted as the scaling dimension of the chiral condensate of the underlying gauge theory at strong coupling, is consistent with the expectation $y\sim 2$ \cite{Aoki:2016wnc,Cohen:1988sq,Ryttov:2017kmx}. We also found that $f^2_\pi/f^2_d=0.086\pm0.015$.

The $\chi^2$ distribution in the full six-dimensional space is relatively flat in the range of $\Delta$ below $\Delta\sim 4.25$. The right panel of Fig.~\ref{Fig:yDeltafits} was obtained by minimizing $\chi^2$ the other five parameters, for each given value of $\Delta$. 
The curve evolves slowly below $4.25$, with values
in the range $3-4$ moderately preferred. The flatness of the curve is due in part to the lesser
accuracy of the lattice data for $M_d^2$. The fit comfortably allows both the 
marginal deformation 
case of $|4-\Delta|\ll1$ as well as the ``mass-deformation'' case of $\Delta = 2$.

%%%%%%%%%%%%%%%%%%%%%%%%sec1%%%%%%%%%%%%%%%%%%%%%%%%%%%%%%%%%%%%%%%%%%%%%%%%%%%%%%
\section{Beyond Leading Order}
\label{sec:BLO}
%%%%%%%%%%%%%%%%%%%%%%%%sec1%%%%%%%%%%%%%%%%%%%%%%%%%%%%%%%%%%%%%%%%%%%%%%%%%%%%%%

In this section, we describe a simple method to determine the dEFT Lagrangian at higher orders in a low energy expansion. We first demonstrate how this method may be applied to determine the form of the leading-order dEFT Lagrangian $\cl_{LO}$ in \mbox{Eq.~(\ref{Eq:L})}. We then apply the method to determine the dEFT Lagrangian at next to leading order (NLO), and comment on the relative size of NLO corrections. Then, employing a spurion analysis, we highlight the symmetry properties of the dEFT, motivating and systematizing the method employed to construct the dEFT Lagrangian.

\subsection{Method for Constructing the Lagrangian}
\label{subsec:method}

In Section~\ref{sec:LO}, we reviewed our leading-order (LO) construction of the dEFT~\cite{Appelquist:2019lgk} and its use to fit the LSD lattice data for the $N_f=8$ gauge theory. The fit made use of the LO Lagrangian in Eqs.~(\ref{Eq:L})--(\ref{Eq:LM}), implemented in the regime $\langle\chi\rangle=F_d\gg f_d$. In this regime, reached by increasing the parameter $m_{\pi}^2$, the scaling laws (Eqs.~(\ref{Eq:Fdvar})--(\ref{Eq:NGBscaling})) insure that the dEFT at LO continues to describe a set of pNGBs and a dilaton, each \emph{relatively} light. The form and estimates of the NLO corrections we present hold when  $F_d\gg f_d$, but we find it simplest to restrict attention to values of $m_{\pi}^2$ for which $F_d$ remains close to $f_d$. Our discussion is then entirely in terms of the lower-case parameters $f_d^2, m_d^2, f_{\pi}^2$ and $m_{\pi}^2$.

The leading order (LO) Lagrangian density in Eq.~(\ref{Eq:L}) consists of terms with one power of the squared masses $m_d^2$, $m_{\pi}^2$, or two derivatives $\partial^2$. The latter generate contributions of order $p^2$ to observables. Each such factor is taken to be small compared to a natural cutoff $\Lambda$, associated with the masses of heavier physical states when $m_\pi^2=0$, which are not included in the dEFT. The dEFT Lagrangian can be described as an expansion in the small, dimensionless combinations $p^2/\Lambda^2$, $m^2_\pi/\Lambda^2$ and $m^2_d/\Lambda^2$, truncated at some given order.

We first observe that the LO terms themselves can be generated in this manner, combined with a single fine tuning.  We begin with a 
"zeroth-order" Lagrangian density of the form

\beq
\cl_0 = \lambda \chi^4\,,\label{Eq:l0}
\eeq
where $\lambda$ is a dimensionless coefficient taken to be of order $\Lambda^2/f_d^2$~\cite{Georgi:1992dw}. We then introduce the following set of dimensionless operators.

\begin{align}
	X_1 & = \left(\frac{\chi}{f_d}\right)^{-2} \;\;\frac{\partial_{\mu}}{\Lambda} \left(\frac{\chi}{f_d}\right)\,, & X_2& = \left(\frac{\chi}{f_d}\right)^{-1}\; \frac{\partial_{\mu}\Sigma}{\Lambda}\,, \nonumber\\
	X_3 & = \frac{m^2_d}{\Lambda^2}\left(\frac{\chi}{f_d}\right)^{\Delta-4}\,, & X_4 &= \frac{m^2_\pi}{\Lambda^2}\left(\frac{\chi}{f_d}\right)^{y-4}\mathbf{1}_{N_f}\,,\label{Eq:bb}
\end{align}
the form of which will be discussed further in the context of a spurion analysis in Section~\ref{subsec:Spu}.

Starting from $\cl_0$, we construct a series expansion in these operators retaining only contributions that are Lorentz invariant. Thus the $X_1$ and $X_2$ operators appear only in pairs.  We also use relations of the type

\beq
1=\frac{1}{N_f}\Tr\left[\Sigma\Sigma^\dagger\right]\, , \qquad \qquad \mathbf{1}_{N_f}=\Sigma\Sigma^\dagger\, ,
\label{Eq:1Tr}
\eeq
and then replace $\Sigma$ (and/or $\Sigma^{\dagger}$) with $X_2$ or $X_4$. Terms in $\cl_{LO}$ (and then $\cl_{NLO}$) are also generated by replacing unity with $X_1$ or $X_3$.

Each of the terms in $\cl_{LO}$, with the important exception of the first term in $V_{\Delta}$, shown in Eq.~(\ref{Eq:Vdelta}), can be generated in this way. The dilaton kinetic term is obtained, up to an $O(1)$ normalization constant, by replacing unity in $\cl_0$ with two factors of $X_1$. The pNGB kinetic term Eq.~(\ref{Eq:Lpi}) is similarly obtained using Eq.~(\ref{Eq:1Tr}) and replacing $\Sigma$ and $\Sigma^{\dagger}$ with $X_2$ and $X_2^{\dagger}$. The pNGB mass term Eq.~(\ref{Eq:LM}) comes, up to an $O(1)$ factor, from the replacement of unity in $\cl_0$ with the first relation in Eq.~(\ref{Eq:1Tr}), the replacement of one factor of $\Sigma$ with $X_4$, and the addition of the Hermitian conjugate. Here we take $f_{\pi}^2$ to be of order $f_d^2/N_f$, a value supported by fits to lattice data for the $N_f=8$ theory.

Turning to the potential $V_{\Delta}$ (Eq.~\ref{Eq:Vdelta}), we first note that the two terms are normalized such that for any value of $\Delta$, the parameters $f_d$ and $m_d^2$ denote the potential minimum and its curvature at the minimum. The limit $\Delta \rightarrow 4$ is smooth. The second term comes from replacing unity in $\cl_0$ with $X_3$, and is small relative to $\cl_0$ for $\Delta$ not close to 4.  Finally, and \emph{very critically}, the first term is already present in $\cl_0$, but in $V_{\Delta}$ with a coefficient suppressed relative to $\Lambda^2/f^2_d$ when $\Delta$ is not close to 4. Thus a single, familiar fine tuning is required to bring this term into line with the others in $\cl_{LO}$. The full expression for $V_{\Delta}$ is small relative to $\cl_0$ for \emph{any} value of $\Delta$.

\subsection{The dEFT at NLO}

We next construct the NLO Lagrangian from $\cl_{LO}$ in Eq.~(\ref{Eq:L}) using the method introduced in Section~\ref{subsec:method}. Each NLO-Lagrangian operator is generated by taking each term from $\cl_{LO}$ and making one replacement with $X_3$ or $X_4$, or two replacements with $X_1$ or $X_2$ from Eq.~(\ref{Eq:bb}). Thus each NLO operator is quadratic in combinations of $m_{\pi}^2$, $m_{d}^2$, and paired derivatives, and also contains a factor $1/\Lambda^2$.  We exclude operators that are parity odd. We also remove operators that are rendered redundant by the equations of motion at LO, or are proportional to total derivatives. By convention, where there are redundancies, we retain the operators with the fewest derivatives.

The NLO Lagrangian contains three kinds of terms, which we group together as follows:

\beq
\cl_{NLO}=\cl_{\pi}+\cl_{p_d}+\cl_{m_d}\,.\label{Eq:lnlo}
\eeq
The first group of terms, $\cl_{\pi}$, contain factors of $m^2_\pi$ and derivatives of the $\Sigma$ field, but no factors of $m^2_d$ or derivatives of the dilaton field:

\begin{align}
	\cl_{\pi}&=l_0\,\,\Tr\left[\partial_\mu\Sigma\partial_\nu\Sigma^\dagger\partial^\mu\Sigma\partial^\nu\Sigma^\dagger\right]+l_1\left(\Tr\left[\partial_\mu\Sigma\partial^\mu\Sigma^\dagger\right]\right)^2+l_2\,\Tr\left[\partial_{\mu}\Sigma\partial_\nu\Sigma^\dagger\right]\Tr\left[\partial^\mu\Sigma\partial^\nu\Sigma^\dagger\right]
	\nonumber \\&+l_3\,\Tr\left[\partial_\mu\Sigma\partial^\mu\Sigma^\dagger\partial_\nu\Sigma\partial^\nu\Sigma^\dagger\right]+l_4\,m^2_\pi\left(\frac{\chi}{f_d}\right)^{y-2}\Tr\left[\partial_\mu\Sigma\partial^\mu\Sigma^\dagger\right]\Tr\left[\Sigma+\Sigma^\dagger\right]\label{Eq:Lnlopi}
	\\&+l_5\,m^2_\pi\left(\frac{\chi}{f_d}\right)^{y-2}\Tr\left[\partial_\mu\Sigma\partial^\mu\Sigma^\dagger\left(\Sigma+\Sigma^\dagger\right)\right]+l_6\,m^4_\pi\left(\frac{\chi}{f_d}\right)^{2y-4}\left(\Tr\left[\Sigma+\Sigma^\dagger\right]\right)^2
	\nonumber
	\\&+l_7\,m^4_\pi\left(\frac{\chi}{f_d}\right)^{2y-4}\left(\Tr\left[\Sigma-\Sigma^\dagger\right]\right)^2+l_8\,m^4_\pi\left(\frac{\chi}{f_d}\right)^{2y-4}\Tr\left[\Sigma^2+\Sigma^{\dagger\,2}\right]+h_2\,m^4_\pi \left(\frac{\chi}{f_d}\right)^{2y-4}.\nonumber
\end{align} 
Using the method of replacements from Section~\ref{subsec:method}, we estimate that each of the dimensionless coefficients $l_i$ and $h_2$ has a size of order $f^2_d/(N_f^t\Lambda^2)$, where the power $t$ counts the number of traces taken in the corresponding operator. It can be seen from Eq.~(\ref{Eq:1Tr}) that each trace comes with a factor of $1/N_f$. The presence of $m_{\pi/d}^2/\Lambda^2$ and $p^2/\Lambda^2$ factors ensures the smallness of these terms relative to the LO Lagrangian.

To illustrate how the terms in Eq.~(\ref{Eq:Lnlopi}) are generated, and how the operator coefficients are estimated using the replacement rules, we consider the first term with coefficient $l_0$ as an example. This operator can be generated starting from the pNGB kinetic term in Eq.~(\ref{Eq:Lpi}) and inserting $\Sigma\Sigma^\dagger$ inside the trace using Eq.~(\ref{Eq:1Tr}). Then, the factor of $\Sigma$ may be replaced with $X_2$, and $\Sigma^\dagger$ with $X_2^\dagger$. Contracting the Lorentz indices one specific way yields the first term in Eq.~(\ref{Eq:Lnlopi}). An independent contraction of Lorentz indices is possible, and is shown as the fourth term with coefficient $l_3$. The order of magnitude of $l_0$ follows directly, without further fine tuning, since the above replacement method multiplies the coefficient $f_{\pi}^2$ in the pNGB kinetic term by $1/\Lambda^2$.
One has $l_0 \sim f_{\pi}^{2}/ \Lambda^2 \sim f_{d}^{2}/ (N_{f} \Lambda^2)$. The natural sizes of the other coefficients in $\cl_{\pi}$ are estimated in a similar way.

The terms in $\cl_{\pi}$ are in one--to--one correspondence with those of an EFT for pNGBs without a dilaton, shown for general $N_f$ in Ref.~\cite{Bijnens:2009qm}. The exponent of the dilaton field in each term is determined by the method of replacement, accounting for the constraints imposed by scale invariance. The form of these terms has been derived before in Refs.~\cite{Li:2016uzn,Cata:2019edh}. When $N_f = 2$ or $3$, trace identities relate some of the operators in Eq.~(\ref{Eq:Lnlopi}), leading to further simplifications. 

The cutoff $\Lambda$ has been estimated in the EFT for pNGBs by calculating the counterterms needed to renormalize the EFT at the one--loop level. The estimates \cite{Georgi:1992dw,Soldate:1989fh} indicate that 

\beq
\Lambda^2 \sim \frac{(4\pi f_\pi)^2}{N_f}\, .\label{Eq:ndaLambda}
\eeq
From this, and the expectation $f^2_\pi\sim f^2_d/N_f$, the order--of--magnitude estimates for coefficients $l_i$ and $h_2$ can be simplified to

\beq
l_i,\,h_i\sim \frac{N_f^{2-t}}{(4\pi)^2}\,.\label{Eq:nda}
\eeq
We have checked that these estimates for the coefficients are consistent with those obtained from the counterterms appearing in Refs.~\cite{Cata:2019edh,Bijnens:2009qm}.

The next group of terms in Eq.~(\ref{Eq:lnlo}) all contain derivatives of the dilaton field, but no factors of $m^2_d$. The non-redundant set of such terms shown below has also been found in Refs.~\cite{Li:2016uzn,Cata:2019edh}

\begin{multline}
	\cl_{p_d} = g_1\,\frac{\left(\partial_\mu\chi\right)^4}{\chi^4}+g_2\,\frac{\left(\partial_\mu\chi\right)^2}{\chi^2}\Tr\left[\partial_\nu\Sigma\partial^\nu\Sigma^\dagger\right]+g_3\,\frac{\partial_{\mu}\chi\partial_\nu\chi}{\chi^2}\Tr\left[\partial^\mu\Sigma\partial^\nu\Sigma^\dagger\right]\\
	+g_4\,m^2_\pi\left(\frac{\chi}{f_d}\right)^{y-2}\frac{\left(\partial_\mu\chi\right)^2}{\chi^2}\Tr\left[\Sigma+\Sigma^\dagger\right]+g_5\,m^2_\pi\left(\frac{\chi}{f_d}\right)^{y-2}\frac{i}{2}\Tr\left[\partial_\mu\Sigma-\partial_\mu\Sigma^\dagger\right]\frac{\partial^\mu\chi}{\chi}\, .\label{Eq:Lpd}
\end{multline}
The $g_i$ coefficients also have a size given by $g_i\sim f^2_d/(N_f^t\Lambda^2)$ and Eq.~(\ref{Eq:nda}).

Finally, we are left with the group of terms that contain factors of $m^2_d$. They are given by:

\begin{multline}
	\cl_{m_d}=
	c_1\,\frac{m^2_d}{(4-\Delta)}\left(1-\frac{4}{\Delta}\left(\frac{f_d}{\chi}\right)^{4-\Delta}\right)\left(\frac{\chi}{f_d}\right)^2\Tr\left[\partial_{\mu}\Sigma\partial^\mu\Sigma^\dagger\right]\\+c_2\,\frac{m^2_dm^2_\pi}{(4-\Delta)}\left(1-\frac{4}{\Delta}\left(\frac{f_d}{\chi}\right)^{4-\Delta}\right)\left(\frac{\chi}{f_d}\right)^y\Tr\left[\Sigma+\Sigma^\dagger\right]\\
	+c_3\,\frac{m^4_d}{(4-\Delta)^2}\left(\frac{\chi}{f_d}\right)^4\left(1-\frac{4}{\Delta}\left(\frac{f_d}{\chi}\right)^{4-\Delta}\right)^2\, .\label{Eq:Lmd}
\end{multline}
Each of the structures above is constructed as a linear combination of an interaction already appearing in $\cl_{LO}$ and the result obtained by replacing a factor of unity from that interaction with $X_3$. The coefficients of the two pieces are chosen so that each structure contains at least one factor of the following expression

\begin{align*}
\frac{m^2_d}{(4-\Delta)}\left(1-\frac{4}{\Delta}\left(\frac{f_d}{\chi}\right)^{4-\Delta}\right)\,.	
\end{align*}	
We choose the two terms and the factor of $1/(4 - \Delta)$ so that it is a smooth, but non--vanishing function of $\chi/f_d$ in the limit $\Delta \rightarrow 4$, as is $\cl_{LO}$. We also include, as a convention, the weighting factor $4/\Delta$, reflecting its defining presence in $\cl_{LO}$\footnote{Other conventions would simply implement small corrections to the parameters within the LO Lagrangian.}. The sizes of the $c_i$ are also given by Eq.~(\ref{Eq:nda}). We are finally left with a total of 18 new operators appearing in the NLO Lagrangian. The full NLO theory therefore has a total of $18+6=24$ free parameters.

\subsection{Spurion Analysis}
\label{subsec:Spu}

By construction, the dEFT is endowed with approximate, spontaneously broken dilatation and internal symmetries. The weak, explicit breaking of these symmetries can be implemented in the dEFT by incorporating spurions into the EFT Lagrangian, and including all the terms allowed by symmetry considerations. In this section, we will employ a spurion analysis to 
show how the replacement rules (introduced in Section~\ref{subsec:method}) emerge and construct the dEFT Lagrangian at LO and NLO.

A spurion is a non-dynamical field that transforms in a given representation of the symmetry group, but then breaks these symmetries once it is assigned a VEV. Since the spurion is not a dynamical field, this introduces explicit rather than spontaneous symmetry breaking. As the appearance of spurions is associated with the presence of the small, dimensionless parameters that control the dEFT expansion, operators in the Lagrangian containing more spurions make contributions to observables that are higher order in our dEFT expansion, and are suppressed once the spurions have been assigned their VEVs. 

As with any EFT, we can build the dEFT employing spurion fields without a detailed description of the underlying gauge theory. We infer the number and symmetry properties of the spurions by comparison with low-energy (lattice) measurements. Once the spurions are chosen, we include all invariant and non-redundant operators at each order in the dEFT expansion. Beyond leading order, observables receive contributions from Feynman diagrams with loops, which can be UV divergent. This procedure will find all the counterterms needed to renormalize the theory, provided that there are no additional sources of symmetry breaking introduced during renormalization.

To allow the dilaton to have mass $m^2_d$ even when the NGBs are massless, we introduce a spurion $\cs(x)$ to break scale invariance without breaking the internal symmetry. Under dilatations $x\rightarrow e^\rho x$, it transforms according to

\beq
\cs(x)\rightarrow e^{\rho(4-\Delta)}\cs(e^\rho x)\,.\label{Eq:strans}
\eeq
It is assigned to a representation of the dilatation symmetry group labeled by scaling dimension $4-\Delta$, which we take to be a free parameter. We measured $\Delta$ for a specific underlying theory by comparing with lattice data, as described in Section~\ref{subsec:Lat}.

We introduce a second spurion $\cm(x)$ to break the internal symmetry (as well as scale invariance) and give the pNGBs a mass. It transforms under dilatations according to the rule

\beq
\cm(x)\rightarrow e^{\rho(4-y)}\cm(e^\rho x)\,,
\eeq
where the scaling dimension $4-y$ is interpreted as a free parameter to be determined from low energy data. We define the spurion field to transform as a conjugate bifundamental field under the unitary transformations $U_{L,R} \in SU(N_{f})_{L,R}$

\beq
\cm(x)\rightarrow U_R \cm(x) U^\dagger_L\,.\label{Eq:mtrans}
\eeq

Under dilatations, the $\chi$ field transforms with scaling dimension one so that $\chi(x)\rightarrow e^\rho\chi(e^\rho x)$, whereas the pNGB field transforms with scaling dimension zero so that $\Sigma(x)\rightarrow\Sigma(e^\rho x)$. Other dimensionful constants which characterize the theory, such as $\Lambda$ or $f_d$ are left unchanged by this transformation.

We construct operators using the spurions, from which the dEFT Lagrangian density is built. We require each operator:
\begin{enumerate}
	\item to be invariant under Lorentz and internal symmetries,
	\item to transform with a scaling dimension of 4 under dilatations. The action will then be invariant under dilatations,
	\item to be polynomial in the spurions and in derivatives.
\end{enumerate}
Crucially, the operators are not required to be polynomial in the dilaton field. We therefore introduce the combination $\chi/f_d$ as a conformal compensator and incorporate it within operators raised to noninteger powers\footnote{The dEFT remains non-singular and well defined even when operators containing negative or noninteger powers of $\chi/f_d$ are incorporated within the Lagrangian, since $\langle\chi\rangle\neq0$.} as necessary to make the operator transform with overall scaling dimension equal to 4 under dilatations. 

To express the LO and NLO Lagrangians in terms of the spurion fields, we employ the replacement method already described, but with the $X_3$ and $X_4$ operators of Eq.~(\ref{Eq:bb}) re-expressed as

\beq
X_3^{sp} = \frac{\cs}{\Lambda^2}\left(\frac{\chi}{f_d}\right)^{\Delta-4}\,,\qquad\qquad X_4^{sp} = \frac{\cm^\dagger}{\Lambda^2}\left(\frac{\chi}{f_d}\right)^{y-4}\,.
\eeq
It can be seen that the operators $X_1$ and $X_3^{sp}$ are invariant under dilatations and internal symmetry transformations. Similarly, it can be seen that the operators $X_2$ and $X_4^{sp}$ are scale invariant but transform in the same way as $\Sigma$ under internal symmetry transformations. It then follows that all valid operators entering $\cl_{NLO}$ (meaning that they meet requirements 1--3) with increasing powers of $\cs$, $\cm$ and more derivatives may be generated from operators entering $\cl_{LO}$ by replacing factors of unity with $X_1$ or $X_3^{sp}$ and factors of $\Sigma$ with $X_2$ or $X_4^{sp}$. Using this replacement scheme and the relation $\Sigma\Sigma^\dagger=\mathbf{1}_{N_f}$, \emph{all} valid operators entering $\cl_{NLO}$ can be generated.

Once we assign fixed values for the spurions so that they no longer transform, as in Eqs.~(\ref{Eq:strans})--(\ref{Eq:mtrans}), the Lagrangian will explicitly break the dilatation and internal symmetries. We set $\cs \rightarrow m^2_d$ and $\cm \rightarrow m_{\pi}^{2}\mathbf{1}_{N_f}$. Setting the $\cm$ spurion proportional to the identity matrix ensures that the diagonal subgroup $SU(N_f)_V\subset SU(N_f)_L\times SU(N_f)_R$ of the internal symmetry is preserved. Giving all $N_f$ Dirac fermions identical masses in an underlying gauge theory also breaks the internal symmetry in this way. Once the spurions have been assigned their fixed values, the $X^{sp}_i$ operators become their corresponding $X_i$ operators, which take the forms introduced in Eq.~(\ref{Eq:bb}).

%%%%%%%%%%%%%%%%%%%%%%%%sec1%%%%%%%%%%%%%%%%%%%%%%%%%%%%%%%%%%%%%%%%%%%%%%%%%%%%%%
\section{Summary and Discussion}
\label{sec:discussion}
%%%%%%%%%%%%%%%%%%%%%%%%sec1%%%%%%%%%%%%%%%%%%%%%%%%%%%%%%%%%%%%%%%%%%%%%%%%%%%%%%

We have reviewed the features and implementation of the dEFT. We deployed it to provide a continuum description of results from lattice studies of an $SU(3)$ gauge theory with $N_f$ Dirac fermions in the fundamental representation, but the dEFT itself is universal, and would describe any theory with the same symmetries and pattern of symmetry breaking at sufficiently low energies.

We first summarized in Section~\ref{sec:LO} the form of the dEFT Lagrangian at leading order (LO) in Eqs.~(\ref{Eq:L})--(\ref{Eq:LM}). It contains a dilaton field associated with the spontaneous breaking of scale invariance in the underlying gauge theory, and a set of pNGB fields associated with the spontaneous breaking of the global internal symmetry of the gauge theory. The spontaneously broken scale invariance is also broken explicitly by a relatively small amount. Similarly, the spontaneously broken internal global symmetry is broken explicitly by a pNGB mass $m_{\pi}^2$ needed to compare dEFT predictions with lattice studies.

We then examined the scaling properties of the dEFT, noting that with $m_{\pi}^2$ increased to values required to describe the lattice data, the explicit breaking of the dilatation and internal symmetries of the dEFT remains small compared to the scale of spontaneous breaking. To this end, the LO Lagrangian was recast in terms of physical (capitalized) quantities in Eqs.~(\ref{Eq:LpiR})--(\ref{Eq:Wexpansion}).

In Section~\ref{sec:BLO}, we examined the structure of the dEFT at next-to-leading (NLO) order. We developed a "replacement method" for identifying dEFT Lagrangian terms at a given order from those at one order lower . We first demonstrated that the LO Lagrangian itself can be generated in this way starting from the "zeroth-order" Lagrangian in \mbox{Eq.~(\ref{Eq:l0})}. As a next step, the method led to the NLO Lagrangian of Eqs.~(\ref{Eq:Lnlopi}), (\ref{Eq:Lpd}) and (\ref{Eq:Lmd}). It is comprised of terms that are generated at the one-loop level and that are naturally suppressed relative to the LO terms. Among the terms of the NLO Lagrangian, some have been discussed in our earlier Ref.~\cite{Appelquist:2019lgk}, and most appear in the recent publications \cite{Li:2016uzn,Cata:2019edh} but some are new. Finally, we motivated the replacement method by showing that it can be derived from a spurion analysis, and checked that its power--counting matches the loop expansion in the dEFT. Composite Higgs models have been built \cite{Appelquist:2020bqj,Appelquist:2022qgl} using dEFT as a foundation, and in this context NLO interactions modify real world Higgs boson properties.

Recent lattice studies of nearly conformal gauge theories provide us with a new opportunity to test longstanding but elusive ideas about spontaneously broken scale invariance in quantum field theory, including the hypothesized dilaton. As a greater variety of more precise lattice data becomes available, it will be important to develop the dEFT further to test the idea. In particular, this requires systematically calculating all contributions at NLO to the observables studied on the lattice, including one--loop diagrams. Furthermore, greater theoretical control over lattice artifacts will be needed. This can be achieved by incorporating the symmetry breaking effects that arise from the lattice discretization within the dEFT itself.

\newpage
%%%%%%%%%%%%%%%%%%%%%%%%%%%%%%%%%%%%%%%%%%%%%%%%%%%%%%%%%%%%%%%%%%%%%%%%%%%%%%
\authorcontributions{

All authors have contributed equally to the conceptualization, formal analysis and writing of this work. All authors have read and agreed to the published version of the manuscript.

}

\funding{
The work of  MP has been supported in part by the STFC 
Consolidated Grants No. ST/P00055X/1 and No. ST/T000813/1.
MP received funding from
the European Research Council (ERC) under the European
Union’s Horizon 2020 research and innovation program
under Grant Agreement No.~813942.

\vspace{0.5cm}

\noindent
{\bf Open Access Statement: } 

For the purpose of open access, the authors have applied a Creative Commons 
Attribution (CC BY) licence  to any Author Accepted Manuscript version arising.

}

\institutionalreview{Not applicable}

\informedconsent{Not applicable}

\dataavailability{Not applicable} 

\acknowledgments{
	
	We would like to thank George Fleming, Pavlos Vranas, and the LSD collaboration for helpful discussions.

}

\conflictsofinterest{The authors declare no conflict of interest.}

%\newpage

%%%%%%%%%%%%%%%%%%%%%%%%%%%%%%%%%%%%%%%%%%
%% Optional
\abbreviations{The following abbreviations are used in this manuscript:\\

\noindent 
\begin{tabular}{@{}ll}
	MDPI & Multidisciplinary Digital Publishing Institute\\
	DOAJ & Directory of Open Access Journals\\
	dEFT & dilaton Effective Field Theory\\
	EFT & Effective Field Theory\\
	LO & Leading Order\\
	NGB & Nambu-Goldstone Boson \\
	NLO & Next-to-Leading Order \\
	pNGB & Pseudo-Nambu-Goldstone Boson \\
	QCD & Quantum Chromodynamics \\
	SM & Standard Model (of particle physics)\\
	VEV & Vacuum Expectation Value \\
\end{tabular}}

%newpage
%\appendixtitles{no} % Leave argument "no" if all appendix headings stay EMPTY (then no dot is printed after "Appendix A"). If the appendix sections contain a heading then change the argument to "yes".
%\appendixstart
%\appendix
%\section{}
%\label{sec:open}
%\subsection{}

%%%
%%% COMMENT THIS LINE IF YOU WANT TO HAVE A TOC
%%
\end{paracol}

%%%%%%%%%%%%%%%%%%%%%%%%appendix1%%%%%%%%%%%%%%%%%%%%%%%%%%%%%%%%%%%%%%%%%%%%%%%%%%
%%%%%%%%%%%%%%%%%%%%%%%%appendix1%%%%%%%%%%%%%%%%%%%%%%%%%%%%%%%%%%%%%%%%%%%%%%%%%%

\end{document}